\documentclass[prl,amssymb,amsmath,showpacs,twocolumn]{revtex4}

\usepackage{graphicx}
\usepackage{amssymb}
\usepackage{amsmath}

\newcommand{\be}{\begin{equation}}
\newcommand{\ee}{\end{equation}}
\newcommand{\bea}{\begin{eqnarray}}
\newcommand{\eea}{\end{eqnarray}}

\begin{document}

\title{Is Quantum Chaos Weaker Than Classical Chaos? }

\author{L.A.~Caron$^{a}$, D.~Huard$^{a}$, H.~Kr\"{o}ger$^{a}$$\footnote{Corresponding author, Email: hkroger@phy.ulaval.ca}$,  
G.~Melkonyan$^{a}$, K.J.M.~ Moriarty$^{b}$ 
and L.P.~Nadeau$^{a}$}

\affiliation{
$^{a}$ {\small\sl D\'{e}partement de Physique, Universit\'{e} Laval, Qu\'{e}bec, Qu\'{e}bec G1K 7P4, Canada} \\ 
$^{b}$ {\small\sl Department of Mathematics, Statistics 
and Computer Science, Dalhousie University, Halifax N.S. B3H 3J5, Canada} 
}

\begin{abstract}
We investigate chaotic behavior in a 2-D Hamiltonian system - oscillators with anharmonic coupling. We compare the classical system with quantum system. Via the quantum action, we construct Poincar\'e sections and compute Lyapunov exponents for the quantum system. We find that the quantum system is globally less chaotic than the classical system. We also observe with increasing energy the distribution of Lyapunov exponts approaching a Gaussian with a strong correlation between its mean value and energy.  
\end{abstract}


\maketitle 
      


\noindent {\it Introduction.}
Quantum chaos has been experimentally observed in irregular 
energy spectra of nuclei,
of atoms perturbed by strong electromagnetic fields \cite{Friedrich},
and in billiard systems \cite{Milner,Dembrowski,Friedman}. Irregular patterns have been found in the wave functions of the quantum mechanical stadium billard \cite{McDonald}, where scars are reminders of classical motion \cite{Heller}.
Recently, dynamical tunneling in atoms between regular islands has been observed. The transition is enhanced by chaos \cite{Raizen,Hensinger}.
Spectra of fully chaotic quantum systems can statistically be described by random matrices \cite{Bohigas}, which corresponds to a level density distribution of Wigner-type, while integrable (nonchaotic) quantum system 
yield a Poissonian distribution. Here we ask: What happens between these two extremes? For example, an hydrogen atom exposed to a weak exterior magnetic field shows a level distribution, which is neither Poissonian nor Wignerian. Can we compare classical chaos with quantum chaos?
And is the quantum system more or less chaotic than the corresponding classical system? 
Also we address the following problem:
An understanding of how classically regular and chaotic phase space is reflected in quantum systems is an open problem. Semiclassical methods of quantisation (EKB, Gutzwiller's trace formula) are not amenable to mixed dynamical systems \cite{Bohigas2}.
Here we suggest a solution using the concept of the quantum action \cite{Q1,Q5}. It parametrizes quantum transition amplitudes in terms of a local renormalized action defined by 
$G(x_{fi},t=T;x_{in},t=0) = \tilde{Z} \exp[ i \tilde{\Sigma}/\hbar ]$,
where $\Sigma$ is a local action - the so called quantum action - evaluated along its classical trajectory going from boundary point $(x_{in},t=0)$ to $(x_{fi},t=T)$. In the limit of large transition time, the quantum action has been proven to exist, giving an exact parametrisation of transition amplitudes \cite{Q4}. The quantum action being local allows to construct a portrait of phase space of a quantum system by applying the tools of nonlinear dynamics to this action \cite{Q3}. In particular, it allows to construct Poincar\'e sections and Lyapunov exponents, in analogy to classical nonlinear dynamics (note that the definition of Lyapunov exponents requires the limit $T \to \infty$). The quantum action has been found useful also to characterize quantum instantons \cite{Q2}.

\begin{table}[h]
\caption{Parameters of classical action vs. quantum action. $v_{22}=0.25$, $T=4.5$.}
\label{tab:ParamAction}
\begin{center}
\begin{tabular}{||c|c|c||}
\hline
Parameter & Class. Action & Quant. Action \\ \hline
$ m $ & 1 & $0.976(8)$ \\ \hline
$ v_{0} $ & 0 & $1.3992(4)$ \\ \hline
$ v_{2} $ & 0.5 & $0.5684(3)$ \\ \hline
$ v_{22} $ & 0.25 & $0.2469(4)$ \\ \hline
$ v_{4} $ & 0 & $-0.00067(7)$ \\ \hline
\end{tabular}
\end{center}
\end{table}

\noindent {\it Model and its quantum action.}
In order to study full chaos the so-called K-system (2-D Hamiltonian, 
potential $V=x^{2}y^{2}$) is widely used.
It is almost globally chaotic, having small islands of stability \cite{Dahlquist}. In order to study mixed dynamics (entangling chaos and regular islands) it is numerically convenient to consider the following classical 
system \cite{Pullen}, 
\bea
\label{eq:ClassAction}
S &=& \int_{0}^{T} dt ~ \frac{1}{2} m (\dot{x}^{2} + \dot{y}^{2}) 
- V(x,y), 
\nonumber \\
V &=& v_{2}(x^{2} + y^{2}) + v_{22} x^{2}y^{2} ~ .
\eea
The parameters of the corresponding quantum action for transition time $T=4.5$ (large compared to dynamical time scale $T_{sc}=1/E_{gr}$) and parameter $v_{22}=0.25$ (which controls chaos) are shown in Tab.[\ref{tab:ParamAction}]. 
They have been obtained by a global fit to transition amplitudes in imaginary time. The bracket gives the estimated error of the fit. 
The numerical results presented below all refer to $v_{22}=0.25$, but qualitatively the same results have been found for $v_{22}=0.05$. 

Before we present our comparison of classical to quantum chaos, we ask what behavior do we expect? In the 1-D quartic potential quantum effects produce a strong positive quadratic term in the quantum action \cite{Q1}.
In the 1-D double well potential, the quantum potential becomes softend, i.e. has lower barrier and closer minima. As a consequence, the instanton solutions of the quantum action are softer than the corresponding classical instantons \cite{Q2}. Are such softening effects also occuring in chaotic phenomena?  

\begin{figure}[h]
\vspace{2pt}
\begin{center}
\includegraphics[bb=213 200 420 579,scale=0.385,angle=0]{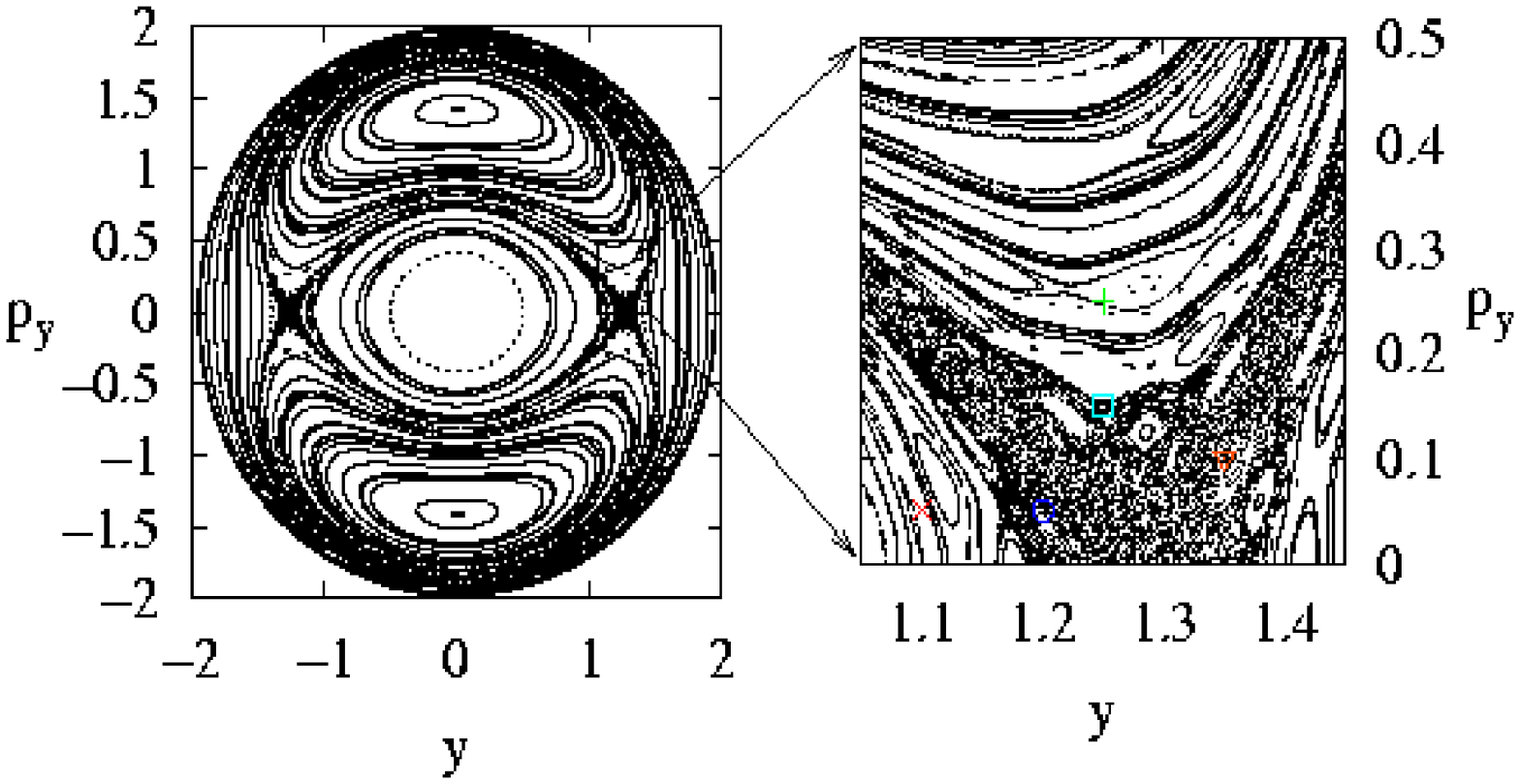}
\vspace{-5pt}
\includegraphics[bb=213 70 599 779,scale=0.365,angle=-90]{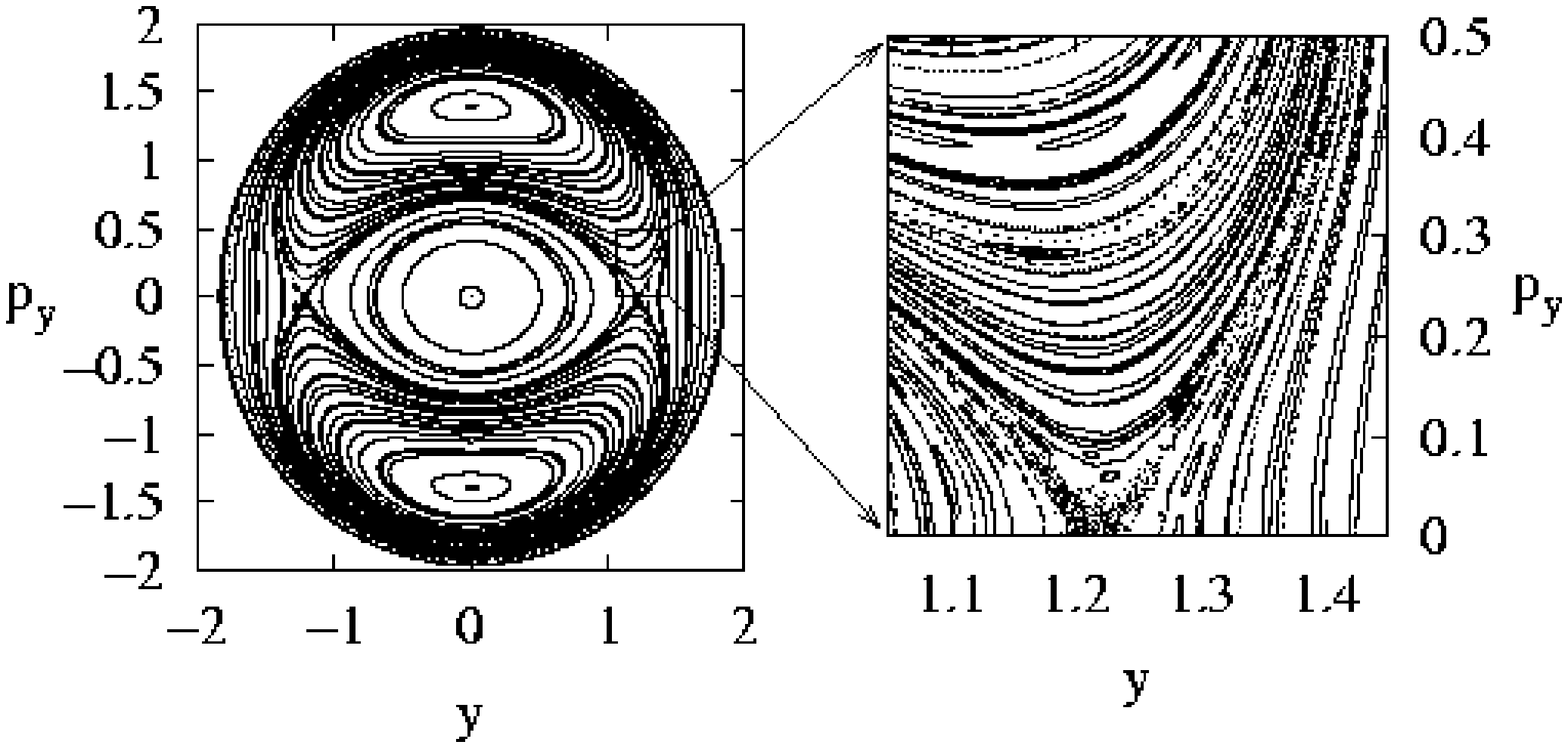}
\end{center}
\caption{Poincare section of classical system (top) and quantum system (bottom). $v_{22}=0.25$. $E=2$.}
\label{Fig_Poin_E2_L25}
\end{figure}

\begin{figure}[tph]
\vspace{9pt}
\begin{center}
\includegraphics[scale=0.4,angle=0]{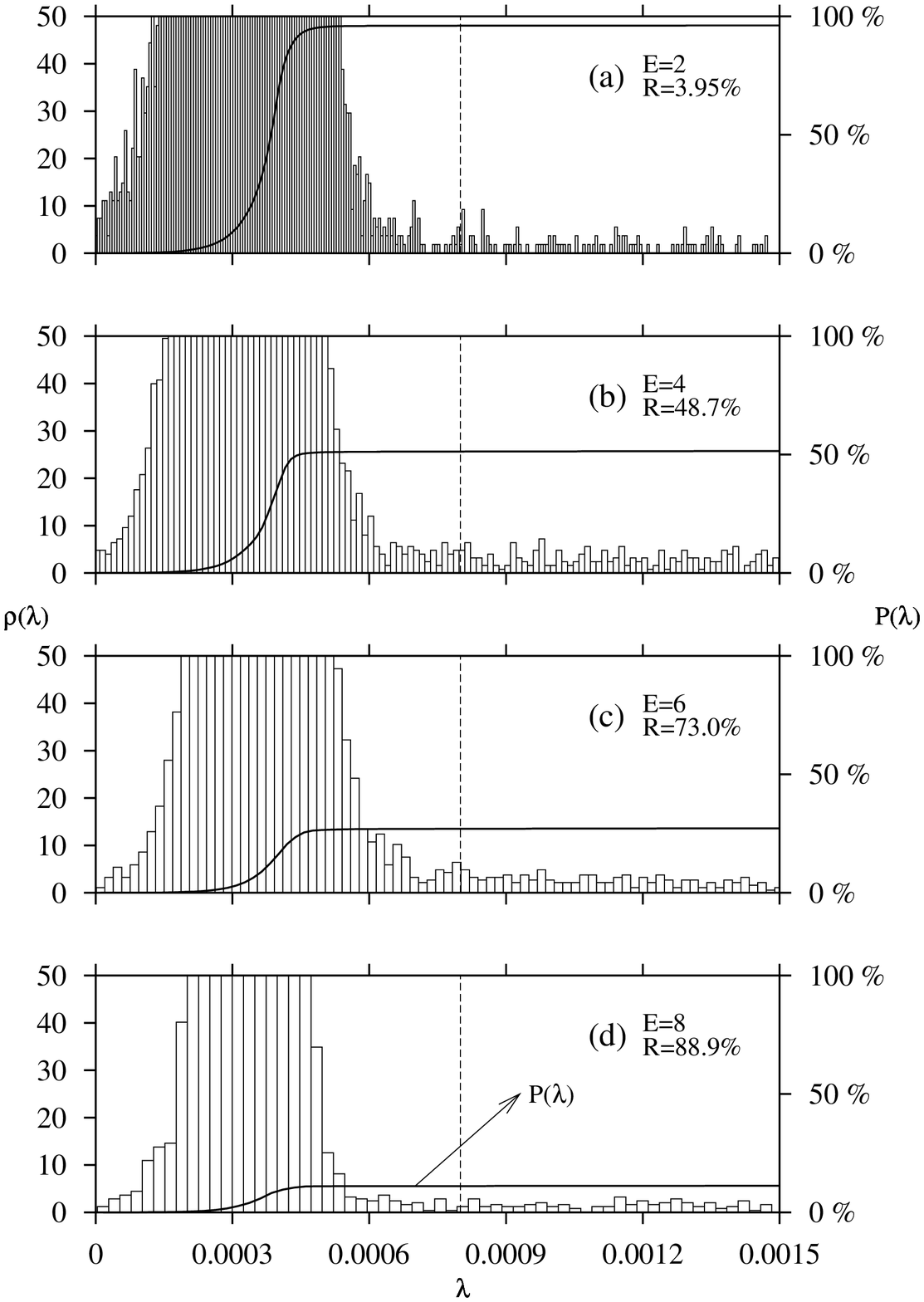}
\includegraphics[scale=0.4,angle=0]{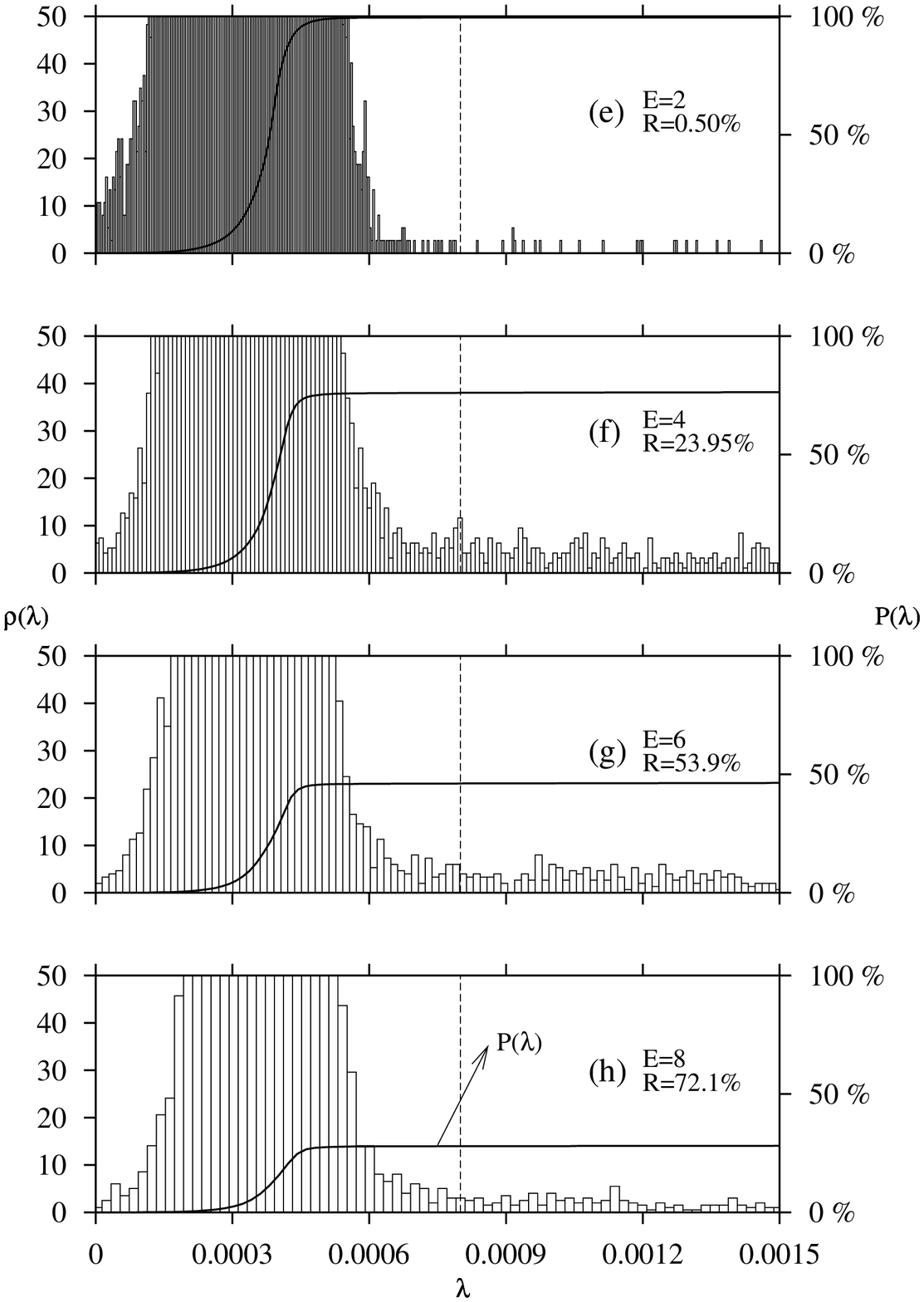}
\end{center}
\caption{Distribution of Lyapunov exponents near $\lambda=0$ for different energies $E$. Classical system (a-d) and quantum system (e-h). $v_{22}=0.25$.}
\label{Fig_Zero_Lyap_4xE_L25}
\end{figure}

\begin{figure}[tph]
\vspace{9pt}
\begin{center}
\includegraphics[scale=0.4,angle=0]{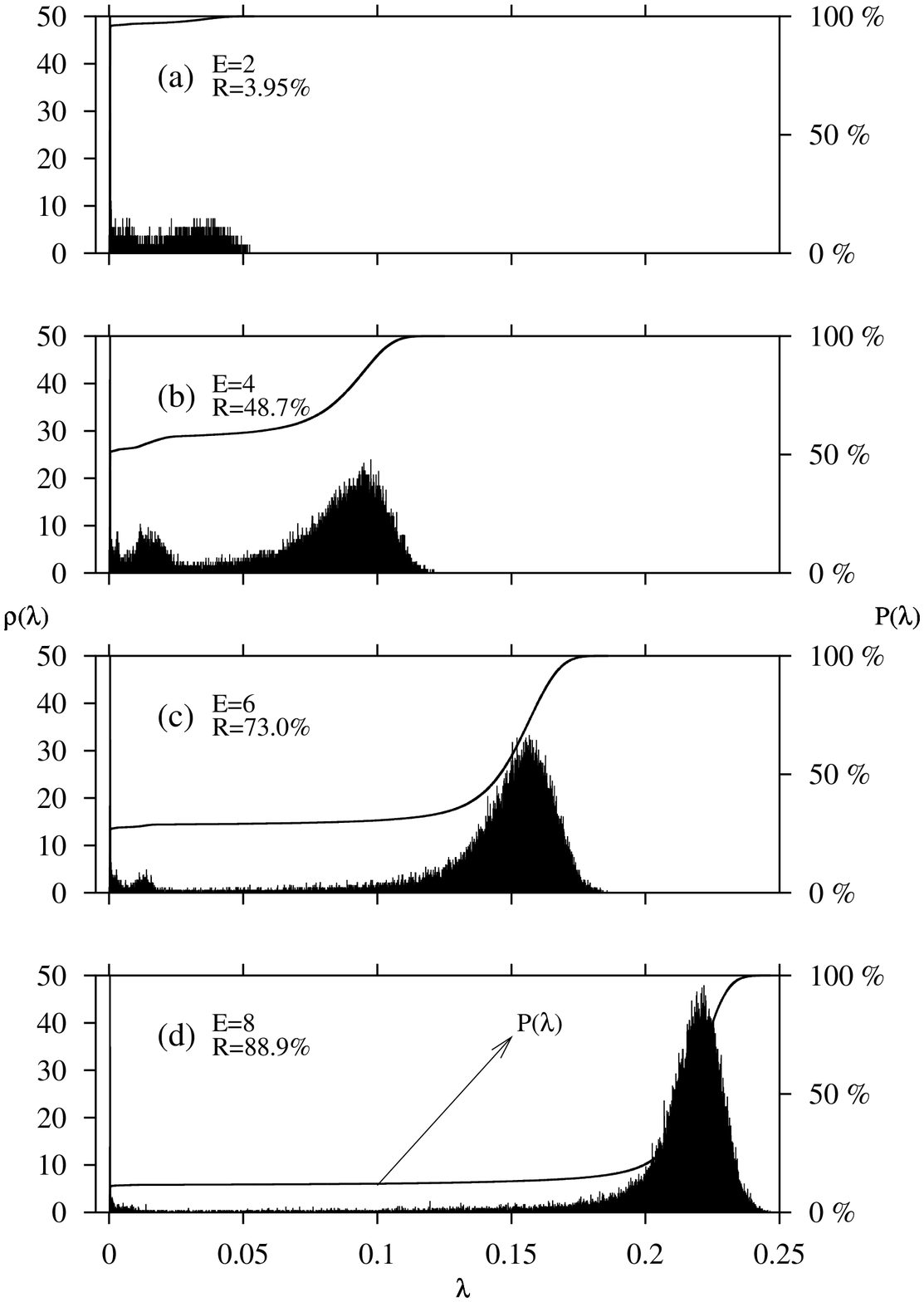}
\includegraphics[scale=0.4,angle=0]{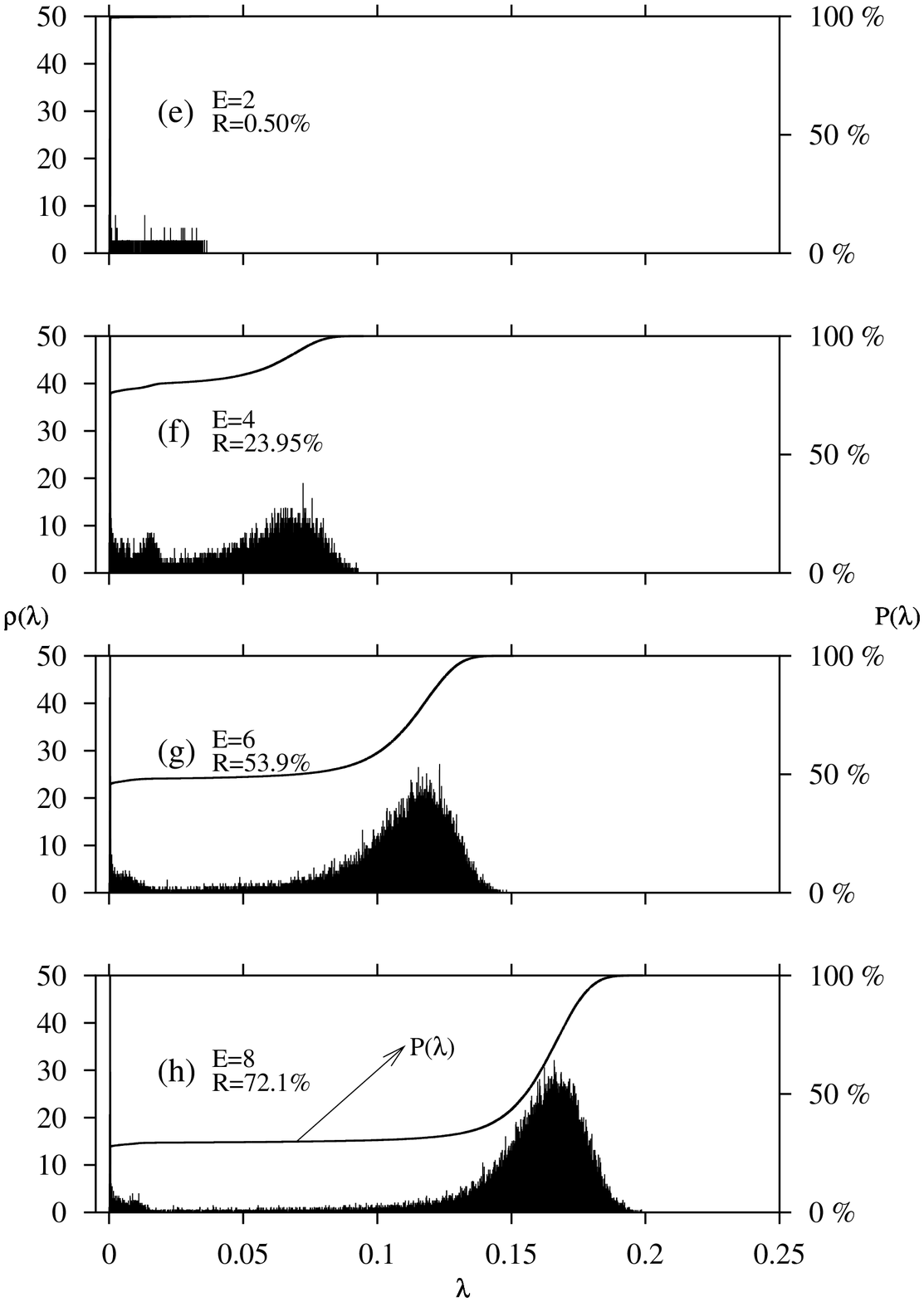}
\end{center}
\caption{Distribution of positive Lyapunov exponents. Classical system (a-d) and quantum system (e-h). $v_{22}=0.25$.}
\label{Fig_Pos_Lyap_4xE_L25}
\end{figure}

\vspace*{5pt}
\noindent {\it Classical chaos vs. quantum chaos.}
Poincar\'e sections for the classical and the quantum system are shown in Fig.[\ref{Fig_Poin_E2_L25}].
One observes an smaller number of chaotic trajectories in the quantum case than in the classical case. This is visible, in particular, in the domain around hyperbolic fixed points (see insert). 
In order to get a quantitative measure for chaotic versus regular phase space, 
we have chosen randomly a large number of initial conditions and computed the finite time Lyapunov exponent for each such trajectory.
The distribution of these Lyapunov exponents in the neighborhood of $\lambda=0$ 
are shown in Fig.[\ref{Fig_Zero_Lyap_4xE_L25}], and those for 
$0 < \lambda < 0.25$ are shown in Fig.[\ref{Fig_Pos_Lyap_4xE_L25}].
Fig.[\ref{Fig_Zero_Lyap_4xE_L25}] shows the distribution of Lyapunov exponents for different energies for the classical and quantum system. While $\lambda=0$ corresponds to regular behavior, $\lambda >0$ indicates to chaotic behavior. For the purpose to distinguish numerically the two regimes, we used this distribution to define some cut-off $\lambda_{c}$ (vertical dashed line in Fig.[\ref{Fig_Zero_Lyap_4xE_L25}]). In order to discriminate numerically $\lambda \lessgtr \lambda_{c}$,  
one has to follow the trajectory for a long time. We used $T_{c}=20000$ to measure $\lambda$. The figure also shows the cumulative distribution (full line), which is a measure for the degree of chaoticity. 
As can be seen this curve is higher for the quantum systen compared to the classical system. 
We denote by $R$ the ratio of chaotic phase space (positive Lyapunov exponent) over total phase space.

The distribution of positive Lyapunov exponents is displayed in Fig.[\ref{Fig_Pos_Lyap_4xE_L25}].
First, one observes a pronounced peak near zero, which has been magnified in 
Fig.[\ref{Fig_Zero_Lyap_4xE_L25}]. Second, with increase of energy, the system becomes more chaotic ($R$ increases). Third, with increasing energy, the distribution of positive Lyapunov exponents seems to approach the shape of a Gaussian. Fourth, the width of the Gaussian (variance) diminishes with increasing energy. Fifth, with increasing energy, a strong linear correlation develops between the expectation value of $\lambda$ and energy $E$, which can be represented by a linear fit $<\lambda> = \lambda^{(0)} + \epsilon E$. All those features hold for both, the classical system and the quantum system. 

\begin{figure}[tph]
\vspace{9pt}
\begin{center}
\includegraphics[scale=0.3,angle=0]{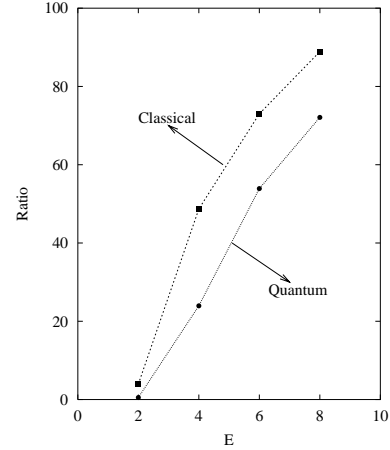}
\end{center}
\caption{Volume of chaotic phase space (positive Lyapunov exponent) over total phase space (lines are guides to the eye). $v_{22}=0.25$.}
\label{Fig_Ratio}
\end{figure}

A quantitative difference between classical and quantum system is apparent in the degree of chaoticity $R$, shown in Figs.[\ref{Fig_Zero_Lyap_4xE_L25},\ref{Fig_Pos_Lyap_4xE_L25}] and plotted in Fig.[\ref{Fig_Ratio}]. 
For all energies, $R$ is smaller in the quantum system.
The same tendency is visible also in the distribution of Lyapunov exponents. For all energies the mean value $<\lambda>$ is smaller in the quantum system (fit parameters are: $\lambda^{(0)}_{cl} = -0.040$, $\epsilon_{cl}=0.033$ and   $\lambda^{(0)}_{qm} = -0.026$, $\epsilon_{qm}=0.024$, i.e. 
$\epsilon_{qm} < \epsilon_{cl}$).
For all energies the classical system turns out to be more chaotic than the quantum system. This is the main result of this work.

How can we understand the observed behavior? 
The distribution of Lyapunov exponents approaching a Gaussian with increasing energy is well known from Hamiltonian system with mixed dynamics \cite{Prasad}. The shape of the curve and its decreasing width can be understood as a consequence of the central limit theorem, which predicts such behavior for an arithmetic mean of $N$ random variables for $N \to \infty$.  
Second, quantum fluctuations which occur in the parameters of the quantum action can be viewed as renormalisation effects of the classical parameters \cite{Q5}. The value of $v_{2}$ (if sole term of potential would render it integrable) is increased, while $v_{22}$ (term which drives the system away from integrability) is decreased (see Tab.[\ref{tab:ParamAction}]. Hence chaos is reduced in the quantum system. Such increase of the quadratic term has been observed in other systems also and seems to occur more generally.

\vspace{0.5cm}
\noindent {\bf Acknowledgements} \\ 
H.K. and K.M. are grateful for support by NSERC Canada. G.M. and D.H. have been supported in part by FCAR Qu\'ebec. 
For discussions and constructive suggestions H.K. is very grateful to A. Okopinska and B. Eckhardt.

\end{document}